# Quantum Measurement without Ontology:
# How Quantum Physical Practice Institutes Objectivity

Richard Healey

## 1. Introduction

What is quantum theory about? What is physically real in the quantum world? These are different questions, and I will offer different answers. Quantum theory is about objective quantum states and Born probabilities, but these are not *physically* real. The Born rule concerns physically real events including measurement outcomes, but quantum theory itself does not describe these objective events. These answers depend on viewing objectivity as instituted by a practice of conformity to epistemic norms, not as accurate representation of physical reality.

Measuring a quantum observable does not generally reveal its value, since not all observables can be consistently supposed to *have* values that measurement might reveal. So, what can we learn from the outcome of a quantum measurement? A system's quantum state (often expressed by a wave-function) does not represent changing values of system observables. Instead, the Born rule specifies the probability of alternative possible events involving a system assigned a quantum state, including possible outcomes of measuring its observables. We can learn about objective quantum states and Born probabilities by measuring observables.

The objective outcomes of our quantum measurements are important because their statistics provide strong reasons to accept quantum theory, and because of their role in quantum metrology. I focus on the latter role in this paper after examining the evidentiary force of our quantum measurement outcomes elsewhere (Healey (2022, 2026)).

Here is a sketch of the rest of this essay, which begins by distinguishing two conceptions of objectivity: veridical representation and conformity to epistemic norms. The first conception faces problems when applied to a fundamental physical theory like quantum theory. These motivate application of the second conception to quantum theory. Before pursuing that application in section 4, section 3 first explains why quantum states, Born probabilities, and measurement outcomes must each be understood to be relative to an appropriate physical situation. This section also distinguishes the absoluteness of a measurement outcome from its objectivity.

Section 5 explains why measurement of quantum observables is important in quantum metrology, even though measurement outcomes do not generally reveal the value of the measured observable. The concluding message is that to understand quantum mechanics one should explain how it is applied so successfully to a physical world it does not itself describe or represent. This message is reinforced by an examination of the successful application of quantum metrology to LIGO to enhance its ability to detect and study the gravitational waves whose observation continues to extend and improve our knowledge of how the physical world is far outside any laboratory.

## 2. Normative Objectivity

Scientific inquiry gives us objective knowledge of the natural world because, or insofar as, this knowledge is arrived at by good (though not infallible) reasoning based on objective data. This data is provided by observation reports and measurement outcomes on which scientists can and do reach widespread agreement. Physics is a basic science and quantum theory has become fundamental to contemporary physics following an unbroken sequence of successful applications during more than a century. Quantum theory has undoubtedly contributed greatly to our objective knowledge of the physical world. But what makes this knowledge objective?

In his seminal work on the origins of objectivity, the philosopher Tyler Burge understood objectivity in terms of accurate representation.

> A central preoccupation of philosophy in the twentieth century was to determine conditions under which it is possible empirically and accurately to represent elements in the physical environment as having specific physical characteristics. Such representation was widely, and I think correctly, taken to constitute a basic type of objectivity. Objectivity in this sense consists in veridical representation of a mind-independent reality. (2009, 285).

His work focused on perceptual representation. But while scientific knowledge would be impossible without perception, most objective knowledge in physics is of things we cannot perceive. We know, for example, that about 100 trillion neutrinos pass through each of our bodies unperceived every second. We know this because we have non-perceptual ways of detecting neutrinos in large and complex instruments located deep underground, and application of quantum theory to nuclear fusion reactions in the sun enables us successfully to predict the resulting neutrino flux and rates of detection, both by these instruments and by human sense organs.

This raises a problem for a notion of objectivity based on veridical representation. We have no better access to mind-independent physical reality than what is provided by applying our best physical theories. We have no independent way of knowing whether those theories represent physical reality veridically, even if application of our best theories does enable us to represent it. This representational sense of objectivity may be applied to perception by humans and more primitive individuals only if one assumes that our best science already provides us with veridical representations of mind-independent reality.

A further problem arises in the case of quantum theory. This theory may be understood in such a way that in application it does not itself represent mind-independent physical reality. QBists maintain, on the contrary, that quantum theory is an extension of decision theory that enables a user to make wiser decisions on how to act in our world.

> For QBism, the quantum formalism is a tool decision-making agents are advised to adopt in light of the peculiar uncertainties we find in our world. (Fuchs (2023))

In a pragmatist view (Healey (2026)) a model of quantum theory is applied not as a representation of mind-independent reality but to offer good advice to a hypothetical localised agent on *how* it may be represented, and on what degree of belief an agent so located should have about each event that is represented as physically possible, given everything such an agent could know about the rest of the physical world.

One can avoid both these problems by adopting an alternative, non-representational, understanding of objectivity. On this understanding, objectivity is a matter of conformity to epistemic norms. Such norms are operative in mathematics. Our knowledge that there are infinitely many prime numbers is objective because that this is so follows, in accordance with norms of deductive reasoning, from what it is to be a prime number. The objectivity of scientific knowledge depends also on reasoning that is good though not deductively valid: Wilfrid Sellars (1953) took such reasoning to involve *material inferences* governed by norms that depend on the content as well as the form of premise and conclusion.

In his inferentialist pragmatism, Robert Brandom (2000) went on to take material inference to play a key role in endowing statements with content. The development of science may alter the norms accepted within a scientific sub-community that govern material inferences involving a statement, thereby modifying its content. Initially, this alteration may affect usage only within that sub-community. But as this development is acknowledged more widely as progressive, the change may be adopted as a norm within the wider scientific community. A norm of deference to experts may eventually affect the content of the statement implicitly acknowledged by all members of the language community. One can see this process at work in changes in the content of statements of simultaneity following the success of Einstein's special theory of relativity, as statements about the absolute simultaneity of distant events are now widely acknowledged to lack objective content, and so should be replaced by significant statements about relative simultaneity.

The objectivity of scientific knowledge depends on norms beyond those governing the significance of statements and inferences between them. Science is a social enterprise governed by defeasible norms of trust. One should accept the data of another scientist as one's own if one has no specific reason not to. Data are provided by reports of observation and outcomes of measurement. Without experience there can be no data. But these are never simply reports of unconceptualized experience (even if there is such a thing), but rest on conscious or unconscious inferences that always involve prior scientific knowledge of the circumstances in which the data is obtained. By accepting another scientist's data, one implicitly acknowledges the objectivity of that data.

Some scientific theories are probabilistic. By fully believing a probabilistic theory one undertakes a commitment to the epistemic norm of matching one's degrees of belief to the probabilities that issue from its correct application, where an application is understood to be correct just in case it conforms to the (typically unwritten) norms implicit in the practice of applying the theory. These norms governing the acceptance and application of a probabilistic theory are what make that theory's probabilities objective, and conformity to them institutes this objectivity in the practice of scientists who adopt it.

## 3. The relativity of Born probabilities, quantum states and measurement outcomes

Section 4 will argue that statements about Born probabilities, quantum states, and measurement outcomes are all objective where these occur in a legitimate application of quantum theory. They are objective even though a Born probability, quantum state, or measurement outcome is not a beable of quantum theory—an element of physical reality represented by that theory. But a complication arises here, because each of the three is also relative to something other than the physical system to which it is applied. So, this section first discusses their relativity and says to what each is relative.

In a sense, according to QBism a quantum state is relative to the observer. But it would be more accurate to say that QBists maintain that when an agent uses quantum theory by assigning a system a quantum state this serves to represent their own doxastic state concerning alternative future experiences they may have. More specifically, it provides an idealised model of that user's personal degrees of belief in their own alternative possible experiences when noting the outcome of their quantum measurement on the system. The idealisation consists in subjecting the user's actual degrees of belief to the internal norm of coherence: degrees of belief that are idealized in this way are often called *credences*. QBists take conformity to the norm of coherence as a requirement of instrumental rationality, imposed on the user's doxastic state to avoid a situation in which they are vulnerable to a Dutch book—a set of bets on possible outcomes of a measurement whose collective outcome is guaranteed to leave them worse off, no matter what outcome they experience.

QBists deny the existence of an additional *external* norm requiring a user's credences to match any objective probabilities prescribed by application of the Born rule to the quantum state of the system. They deny that the quantum state assigned by the user is subject to any external norm capable of rendering it objective. Indeed, no quantum state is even represented in their preferred formulation of the Born rule. They reformulate this rule as another "empirically justified" *internal* norm relating a user's credences concerning their possible experiences of outcomes following alternative possible measurement actions they may perform on a system.

A type of measurement may be informationally complete, in the sense that the agent believes that very many repetitions of that type of measurement on systems the agent assigns the same quantum state would yield frequencies of experienced outcomes sufficient for the agent confidently to assign that state. Their reformulated Born rule relates a user's credences concerning outcomes of an arbitrary measurement to the user's credences concerning the outcomes of an informationally-complete measurement. QBists like to express this relation in terms of a class of informationally-complete measurements dubbed *symmetric* because in this case it takes a simpler form, for a system whose possible quantum states may be represented in a Hilbert space of natural number dimension $n$.

Some form of the Born rule is applied to an enormous range of different kinds of systems—in atomic, nuclear, elementary particle and condensed matter physics; within non-relativistic and relativistic quantum mechanics; within Lagrangian, algebraic and axiomatic quantum field theory; and in rival approaches to a quantum theory of gravity. If the QBists' reformulation of this rule were equivalent to the original Born rule, one would expect to see users of quantum theory acting in conformity to this norm in all these applications. But they do not. Quantum theory can be successfully applied to systems whose quantum states are represented on a Hilbert space of countably infinite dimension, and also to systems whose states are represented on a non-separable space of states on a von Neumann algebra of observables. A form of the Born rule applies to quantum states represented in all these spaces.

The QBist reformulation of the rule makes sense only in the special case of systems associated with finite-dimensional Hilbert spaces.

Even though the QBist reformulation of the Born rule is limited in application and rarely (if ever) taught to students of quantum theory, physicists may well acknowledge it as a norm governing some applications of quantum theory after it has been explained to them. Its normative status is then inherited from the more basic norm expressed by the Born rule in the general form in which it connects the quantum states to which it is applied to the quantum probabilities it yields. Section 4 argues that acknowledgement of this general form of the Born rule helps to make quantum state assignments objective.

Proponents of relational quantum mechanics also maintain that a quantum state is relative to the observer, but they do not require an observer to be an agent using quantum mechanics, or even something capable of having doxastic states.

> By using the word "observer" I do not make any reference to conscious, animate, or computing, or in any other manner special, system. I use the word "observer" in the sense in which it is conventionally used in Galilean relativity when we say that an object has a velocity "with respect to a certain observer". The observer can be any physical object having a definite state of motion. For instance, I say that my hand moves at a velocity v with respect to the lamp on my table. Velocity is a relational notion (in Galilean as well as in special relativistic physics), and thus it is always (explicitly or implicitly) referred to something; it is traditional to denote this something as the observer, but it is important in the following discussion to keep in mind that the observer can be a table lamp. (Rovelli (1996))

According to relational quantum mechanics, a quantum system has a quantum state only relative to another quantum system with which it has undergone a quantum interaction modelled by unitary evolution of their joint quantum state. Measurement of an observable on a quantum system involves such an interaction with a second system, and produces a physical outcome recorded in the (relative) state of that system (and possibly also in the relative state of the first system).

Advocates of relational quantum mechanics have been challenged to produce clear and defensible specifications of the kind of interaction required for measurement of a given observable and of exactly when it has led to an outcome (Muciño, Okon and Sudarsky (2021)). They have also faced a difficulty concerning the possibility of intersubjective agreement among the recorded outcomes of multiple measurement-type interactions on a system, and of measurements by each observer system of the outcomes of the measurements of the other observers. The proposal to secure such intersubjectivity by adding the assumption of cross-perspective links (Adlam and Rovelli (2023)) would not restore objectivity of measurement outcomes even if it facilitated intersubjective agreement. In a recent paper Riedel (forthcoming) argues that adding such an assumption to relational quantum mechanics even rules out the possibility of absolute facts about *relative* measurement outcomes.

Problems that arise when one takes quantum state assignments, Born probabilities and measurement outcomes to be relative to quantum systems or to agents can be solved by recognizing that all of them are relative not to an agent or quantum system but to a physical situation of one kind or another. This recognition is a key feature of a pragmatist view (Healey (2012, 2026)). Before defending the objectivity of Born probabilities, quantum states and measurement outcomes from this pragmatist viewpoint I first specify the different kinds of physical situation to which they should be relativized.

In this pragmatist view, a quantum state is assigned relative to an *agent-situation*. An agent-situation is not an agent. It includes a localized spatio-temporal region that might or might not be occupied by a spatially-localized agent during some interval. To further specify an agent-situation, one must describe not only that region, but also physical conditions in its neighbourhood, sufficient to determine what physical processes are in place permitting any agent who might be located there to acquire knowledge of physical conditions elsewhere. This neighbourhood is confined to the causal past of the region, since an agent located in a space-time region can know nothing about what lies outside its causal past, no matter what processes occur there. But if no physical processes are in place permitting any agent located in a space-time region to acquire knowledge about an event in its causal past, then that event also remains epistemically inaccessible to that agent.

In this view a measurement outcome is a knowledge claim, about a physical event involving a quantum system $S$ that occurs in a space-time region following a unitary interaction between $S$ and a second quantum system $A$ one might call an "apparatus". The event consists in $A$ acquiring a property $P=p_i$. One can apply a model of decoherence to assess the suitability of the interaction for measuring an observable $M$. The interaction is suitable for measuring $M$ if the reduced state of the apparatus system in the model of these systems in their environment rapidly and robustly decoheres in a "pointer basis" of eigenstates of its "pointer observable" $P$. This justifies one in taking the value $p_i$ of the pointer observable to indicate the corresponding outcome $M\varepsilon\Delta_i$ of the measurement (where $\Delta_i$ is a Borel set of real numbers), even though in the model decoherence is neither perfect nor final. Inferentialist pragmatism accounts for the meaning of the *magnitude claim* $M\varepsilon\Delta_j$, by tracing it to the reliability of material inferences connecting this claim to other meaningful statements. That is why a measurement outcome is relative to the *decoherence environment* in whose context each claim $M\varepsilon\Delta_j$ that here expresses the possible outcome of a measurement of $M$ is meaningful, and either true or false; and exactly one such claim $M\varepsilon\Delta_i$ is true.

What counts as a measurement outcome depends on the quantum state whose decoherence in a model underlies its significance and truth-value. Since that quantum state is itself relative to an agent-situation, so too is each possible measurement outcome. But the relation between agent-situation and decoherence environment is many-one: many alternative agent-situations correspond to the same decoherence environment. So, agents who happen to occupy somewhat different agent-situations should typically agree on the outcome of measuring an observable (relative to their shared decoherence environment). However, the situations of Wigner while outside, and his friend while inside her physically isolated laboratory would be so different that the friend's measurement has an outcome relative to the decoherence environment inside her laboratory but no outcome relative to the decoherence environment outside.

## 4. The objectivity of Born probabilities, quantum states and measurement outcomes

There is a pragmatist view in which Born probabilities, quantum states and measurement outcomes are all objective, even though none of them are part of the ontology of quantum theory. They are not what Bell called ‘beables’ of the theory—elements of physical reality if quantum theory veridically represents them. In this view, quantum theory itself has no such physical ontology, which is a reason to call the view desert pragmatism (Healey (2022)). A competent measurement of a quantum observable gives rise to a physical event that one may be justified in claiming as its outcome. But that event is not a quantum beable, and nor is the probability assigned to it by a legitimate application of the Born rule. Indeed, unlike the event claimed as a measurement outcome, Born probabilities are not even physical things, and nor is the quantum state that yields them in an application of the Born rule.

Born probabilities, quantum states and events claimed as measurement outcomes all exist, but none of them is an element of mind-independent physical reality veridically represented by quantum theory. An event claimed as the outcome of measuring a quantum observable is physical. But it is not represented by quantum theory. Quantum theory is applied only to make sure one can meaningfully claim it as the measurement outcome and to specify its probability. Though represented in this application of the Born rule, that probability is not a beable of quantum theory since it is not an element of physical reality: it is neither a physical propensity nor a relative frequency. Quantum theory does represent the quantum state assigned to a system—veridically if this assignment is correct. But even when real, a quantum state is not an element of *physical* reality: it has no spatio-temporal location, carries no energy or momentum, and has neither causes nor effects. Nor does the quantum state of a system represent any of its physical properties, through the eigenstate-eigenvalue link or otherwise. So, it is misleading to call an equation describing (or “governing”) the evolution of a system’s quantum state a dynamical law.

In defending the view that quantum probabilities, states and measurement outcomes are objective despite being relative I will begin with quantum probabilities as they figure in the Born rule. Quantum theory is popularly believed to have shown that the world is indeterministic—that the past determines only the objective chance of each of various possible future continuations. QBists follow De Finetti and others in denying the existence of any kind of objective chance, taking quantum and all other probabilities to represent just the personal degrees of belief of an instrumentally rational agent. Temporarily assuming the role of a subjectivist, the philosopher David Lewis (1980) nevertheless acknowledged the concept of objective chance in providing his influential subjectivist’s guide to this concept as, he thought, it figured (only) in fundamental physics including quantum theory.

Lewis was right to acknowledge objective as well as subjective concepts of probability: he called these chance and credence, respectively. Lewis’s Principal Principle was a valuable attempt to say how these concepts are related. But improvements to his Principal Principle show both why he was wrong to think that objective probability figures only in fundamental physics, and also why objective probability is *always* relative to an agent-situation, in quantum theory and elsewhere.

The idea behind Lewis’s Principal Principle is that an agent recognizes a concept of objective chance by conforming their degrees of belief to what they believe that chance to be. If there were no such thing as chance (objective probability), then this belief about its value would be mistaken, and so they would be wrong to recognize the concept. But Lewis believed that fundamental physics was committed to the concept of objective probability by the successful deployment of this concept in the theory of radioactive decay processes. Since that theory is quantum, he implicitly acknowledged the objectivity of quantum theory's Born probabilities. In Lewis's view, uses of probability in the rest of science, including in non-fundamental theories like classical statistical mechanics, involved no commitment to objective chance: they could all be understood as applications of the subjective notion of credence.

In stating his Principal Principle, Lewis helped himself to an absolute notion of time and a distinction between admissible and inadmissible information. Both the objective chance of an event and the distinction between admissible and inadmissible information about its occurrence are indexed to a time. They may vary as time passes, since information is admissible at a time only if possessing this information would have no effect on an agent's degree of belief about the chance of that event at that time. An agent who believes that a future event may or may not occur but takes its chance of occurring to be ½ should change their belief about its chance only if they were to receive further information about whether it will actually happen. Commitment to a non-trivial concept of objective chance here depends on considering such foreknowledge to be inadmissible.

A notion of absolute time plays no role in contemporary space-time theories. Instead, these theories take the light-cone at each space-time point to divide space-time locally into that point's causal past, causal future, and a space-like separated region often called the absolute elsewhere. The Principal Principle must be reformulated to be applicable to a concept of objective chance suited to such a relativistic space-time. One should think of objective chance as relativized not to a moment of time but to a space-time point. The distinction between admissible and inadmissible information must also be modified. The function of this distinction in Lewis's statement of his Principal Principle was to count information as admissible at a time only if anyone applying that principle then could (in principle) obtain it by means of physically possible processes. Assuming that information cannot be transferred faster than the velocity of light in a vacuum, one located at a space-time point should count as admissible only information about events in or on its back light-cone, while information is admissible at a space-time region only about events in its causal past.

A notion of objective probability (Lewis's chance) that functions according to a Principal Principle modified to make sense in a relativistic space-time is therefore automatically specified relative to a space-time location, in just the way that Lewis's non-relativistic concept of chance was specified relative to an absolute time. An agent acknowledging such a concept must accept that the objective probability of an event is relative to a space-time region they take to mark the distinction between admissible and inadmissible information.

In Lewis's original non-relativistic formulation, the distinction implicitly appeals to the whole of space at a moment of absolute time: in its relativistic reformulation the distinction appeals to a point of space-time. In each case this represents an idealization of the physical location of a possible situated agent then contemplating application of that principle to form credences about the objective probability of an event on the basis of all admissible information. More realistically, that agent's situation would be represented by a space-time region—a small region of space throughout an interval of absolute time—part of a "thick time-slice" marking the period relative to which a hypothetical agent present then might apply Lewis's original Principle, or a compact region of a relativistic space-time relative to which an agent might apply the modified Principle. That is why a concept of objective probability makes sense in a relativistic space-time only relative to an agent-situation whose specification includes a space-time region an agent might occupy for a while, whether or not any agent occupies that region.

An epistemic agent occupying a compact space-time region of a relativistic space-time may face a significant barrier that blocks access to information about the contents of their causal past. Each of us is confronted by such a barrier in making a typical application of classical statistical mechanics. Even though potentially relevant information about the earlier positions and momenta of all the molecules of a gas and its container is present in the causal past of anyone making the application, they then have no way of accessing this vast amount of detailed information and applying classical mechanics to predict the future behaviour of the gas. They correctly judge such information to be inadmissible when applying a corresponding concept of objective probability when applying classical statistical mechanics. This is just one example of a use of objective probability in non-quantum physics. Other examples occur in other areas of science, including evolutionary biology as well as weather and climate forecasting.

Armed with this understanding of what is involved in acknowledging a concept of objective probability, one can now raise the question of whether the practice of quantum physicists shows that they *do* acknowledge the objectivity of Born probabilities. It does. By accepting quantum theory, one treats the Born rule linking quantum states to Born probabilities as a norm governing one's degrees of belief, and consequently one's actions. Acting in conformity to this norm is one way in which the objectivity of Born probabilities is instituted in the practice of physics.

Physics education institutes another way. Students learning physics are taught the Born rule as a norm of quantum theory. An unusually independent-minded student may be reluctant to wholly accept the theory being taught. Perhaps aware of radical theory changes in the history of physics, he or she may recognize that all scientific knowledge remains fallible. The credences of such a student may not wholly conform to the norm being taught, so their actions may deviate from those of their fellows. But quantum theory has been so successfully applied to predict, explain and control natural phenomena that any failure to conform one's credences to Born probabilities will be subject to severe criticism by members of the scientific community. Such enforcement activity is another element of physical practice that institutes the normative character of the Born rule.

Since issuing in Born probabilities is the primary function of quantum state assignment, the relativity of Born probabilities to agent-situation implies that assignment of a quantum state to a system is also relative to agent-situation. But the objectivity of quantum state assignments follows from the normative status of the Born rule only if quantum state assignments are themselves objective. QBists deny that a quantum state assignment is objective; so, they can consistently accept the Born rule (in a formulation in which it links a quantum state to the Born probabilities it yields) as a norm of quantum theory, while denying that Born probabilities are objective.

There are aspects of quantum physical practice that nevertheless institute the objectivity of quantum state assignments. Theoreticians speak of the operation of preparing the quantum state of a system, and experimenters devote considerable effort to executing such operations and verifying that they have successfully prepared the desired quantum state. The successful operation of a quantum computer depends on correctly setting its initial state to begin the computation. Physical gates that implement the intended transformations from this state must then perform as intended for the computer to work correctly. The operation of these physical qubit elements is subjected to rigorous tests of their error rates and designed to minimize these. The theory and practice of error correction is based on coping with such errors by introducing additional physical qubits to produce logical qubits with lower error rates. The concept of an error is applicable only as deviation from correctness, here understood as conformity to an operative norm.

The controlled preparation and transformation of quantum states is central to applications of quantum theory in many areas of contemporary physics, including quantum optics and condensed matter physics. In many applications, these states are used to observe and measure other things; in interaction-free quantum imaging, in gravitational-wave observatories, to construct accurate clocks and gravimeters. Each such application contributes to physical practice governed by a norm that one obeys only by assigning the correct quantum state to a system (relative to the agent situation of anyone who might apply the Born rule to determine the objective probabilities of alternative events involving that system). The correct quantum state is real, but it is not (and does not represent) an intrinsic physical property of that system. It is real just because relative to the relevant agent-situation a statement assigning the system that state is true.

A final example of the institution of objective quantum states through physical practice is provided by the award of a 1998 Nobel prize for physics to Robert B. Laughlin for contributing to our understanding of the fractional quantum Hall effect. In his seminal publication, Laughlin hypothesized that what has come to be known as the Laughlin wave-function correctly represents the quantum state of a quantum system—the ground state of a two-dimensional electron gas in a strong magnetic field. He used this hypothesis to explain the observation of the fractional quantum Hall effect, and he predicted the existence of additional states as well as quasiparticle excitations with fractional electric charge , both of which were later experimentally observed. This pattern of reasoning in support of his hypothesis about the quantum state of the system is typical of that used to establish any empirical hypothesis as an item of objective scientific knowledge. The Nobel committee took its conclusion that the system's state is correctly represented by the Laughlin wave-function as an outstanding example of objective knowledge by awarding him a Nobel prize.

Quantum states and the consequent Born probabilities are objective not because of what they represent but because they are treated as norms of quantum theory. By wholly accepting quantum theory an agent is committed to forming credences in meaningful claims about the possible values of observables on a quantum system that match the Born probabilities of these claims. Such a claim acquires the necessary meaning in a suitable decoherence environment, such as that engineered in an experiment designed to measure the value of that observable. But the Born rule may be legitimately applied to assign probabilities to events that are not measurement outcomes. It may be applied in the context of a decoherence environment that occurs without the help of any agent and outside any laboratory. It may be applied to events that occur in the interior of a neutron star, or to events one takes to occur in a merely possible world containing no agents. Any agent making the application occupies an agent-situation in this world, of course. But the quantum state used in the application may be relative to some quite different agent-situation no agent occupies.

A decoherence environment for an observable relative to one agent-situation need not be a decoherence environment for that observable relative to a different agent-situation. Wigner's friend occupies an agent-situation relative to which there is a decoherence environment for the observable she correctly takes herself to measure in her laboratory. Wigner occupies an agent-situation outside the laboratory relative to which there is no decoherence environment for that observable, so he correctly concludes that she has performed no measurement. There is no conflict between them, because whether a measurement has an outcome is relative to decoherence environment, and indirectly to agent-situation. But in less recherché scenarios, for a considerable variation in agent-situation there is no corresponding variation in decoherence environment. In ordinary applications of quantum theory relative to different agent-situations, the decoherence environment is the same relative to each of those agent-situations. So, an agent applying quantum theory relative to any of these different agent-situations will reach the same conclusions, about the occurrence and outcome of a quantum measurement in that decoherence environment.

Unlike quantum states and probabilities, the events that occur relative to a decoherence environment are physical. These include events whose occurrence physicists claim as the outcome of a measurement of a quantum observable. There is a serious issue as to whether such a claim can contribute to objective scientific knowledge if there is no absolute fact about the outcome of the measurement. It is not only this pragmatist view of quantum theory that must face this issue.

Recent no-go proofs set in an extension of the scenario of Wigner's friend (an EWFS) use the assumption that there is an absolute fact about the outcome of each of a set of quantum measurements performed in that scenario to derive a contradiction, given other very plausible assumptions about any theory capable of predicting how these outcomes are correlated (Schmid, Ying, and Leifer (2023) review several of these proofs). Desert pragmatism addresses this issue by first distinguishing the absoluteness of an event from its objectivity. This permits one to explain how and why a measurement with an outcome relative to one agent-situation but no outcome relative to another agent-situation may nevertheless be objective in the sense that matters for science. Once more, objectivity in this sense is not to be understood as veridical representation, but in terms of conformity to norms instituted by good scientific practice.

Three norms of scientific practice institute the objectivity of quantum measurement outcomes. I follow Tal (2017, 34) in taking a measurement outcome to be a knowledge claim attributing one or a range of values to the object or event being measured, a claim that is inferred from one or more instrument indications along with relevant background knowledge. An instrument indication is a property of a measuring instrument in its final state after the measurement process is complete.

There is a norm of trust, a norm of personal observation, and a norm of verification. Closely related norms have application outside of scientific activity. One can find such applications in Shakespeare's play Macbeth, as well as in Ronald Reagan's approach to nuclear disarmament discussions with Mikhail Gorbachev. Doubting the objective presence of the dagger he saw, Macbeth tried to grasp it: to prove that Banquo's ghost occupied his own place at the table, he pleaded his guests to see for themselves. Reagan frequently repeated the phrase "trust but verify", the translation of a Russian proverb. Here is a statement of each of three norms governing scientific practice.

*Trust*

An instrument indication or measurement outcome claimed by one scientist should be accepted by other scientists provided that there is no specific reason for doubting the claim.

*Personal Observation*

A scientist who personally observes the indication of an instrument should accept the claim that the instrument has this property, provided that (s)he has no specific reason for doubting this claim.

*Verification*

If an observable is measured on a system by each of several independent methods then their outcomes should be accepted only if they agree.

Note that failure of agreement in an attempted verification would provide one kind of specific reason for the doubt mentioned in the proviso included in each of the first two norms.

No-go proofs give strong reasons to believe that, even when it justifies a claim about the measurement's outcome, an instrument indication after careful execution of a measurement procedure for a quantum observable on a system does not justify the claim that this was the value of that observable.[1] Agreement between instrument indications in *successive* independent measurement procedures of a quantum observable performed on a single system provides no better reason to believe that these procedures reveal the value of the observable on that system. For agreement among independent methods to provide a stronger reason to accept a measurement outcome, each method must be employed as part of a larger physical procedure that occurs on a single occasion.

Quantum Darwinism (Ollivier, Poulin and Zurek (2004), Zurek (2009)) models such independent measurements as interactions with separate parts of a system's wider environment, following an initial interaction between the system and one or more systems in its immediate environment. In a desert pragmatist view, a user of quantum theory applies these models of decoherence quite differently from the way they are applied by Zurek and his collaborators—not to understand the emergence of objective quasi-classical reality from an underlying "subjective" quantum state, but to understand the advice they offer to a hypothetical situated user on how, and how far, physical events (including those claimed as measurement outcomes or as properties acquired by instruments) may be considered to be objective.

The three stated norms governing good scientific practice all focus on the practice of measurement. Even *Personal Observation* used the word 'instrument', while the others spoke of measurement outcomes. In deploying a model of environmental decoherence, it is important for a desert pragmatist not to rely on such terminology, since these models are taken to apply to a class of possible interactions, some of which occur in no laboratory, involve no devices constructed by experimenters, and in a world in which there are no agents. They should be replaced by the terms ‘agent-situation’, ‘decoherence environment’, and ‘magnitude claim’. But this does not make application of models of environmental decoherence irrelevant to the three norms.

[1] These include Kochen and Specker (1967) and Fine (1982).

Application of a quantum model of environmental decoherence to an instrument relative to an agent-situation *A* occupied by an agent observing that instrument is what would certify that agent in holding a claim about its instrument indication meaningful enough to count as either true, or on the other hand false, relative to *A*. Applying the second norm, the agent could then appeal to their personal observation as a reason justifying their claim about the instrument’s indication relative to *A*. An agent not occupying situation *A* can appeal to the first norm as a reason justifying the same claim relative to *A*.

A measurement outcome is justified by an instrument indication only in the light of a model of the measurement process. For a desert pragmatist, in the case of a single measurement of an observable on a system, this will include a quantum model of environmental decoherence. But that model is not used to represent the physical process or processes involved. So, how can its application justify any claimed outcome of the measurement?

Application of a model of environmental decoherence can certify that a claim about the value of a pointer observable on a system $S_{n+1}$ has enough content to be assigned a truth-value relative to a decoherence environment following an interaction with a system $S_n$. But the model cannot represent that truth-value: only an agent in an agent-situation relative to which that is such a decoherence environment could observe the pointer observable and say whether the claim is true or false. These conditions apply equally for subsequent interactions between $S_{n+1}$ and a further system $S_{n+2}$. ($n=1,2,...$) until one reaches a system *I* that a scientist can personally observe, at which point that scientist can say what the pointer observable of *I* indicates. Other scientists should then endorse this claim, abiding by the norm of *Trust*.

A sceptic might object that since no agent occupied an agent-situation from which to personally observe *S* or any $S_n$ prior to *I*, one can consistently imagine that the outcome of the measurement on S is uncorrelated with the value of the pointer observable of each prior system in the sequence, and with that indicated by the pointer observable on *I*. If that were so, then observation of *I* would not provide any justification for a knowledge claim as to the outcome of the measurement on *S*.

But a scientist should dismiss this imagined scenario as merely a sceptical possibility that does not offer a specific reason for doubting that the instrument indication of *I* justifies that knowledge claim. That claim may receive further support from independent observations by the same or different scientists of the indications of other instruments *J*, *K*, *L*, … , each of which is connected to the original system *S* by its own chain of interactions modelled by quantum decoherence. Quantum Darwinism emphasizes the importance of this possibility by showing that multiple independent records of the outcome of a quantum measurement on *S* are available in different parts of its environment, each of which may generate a personally observable indication in a different instrument. Even personal observations by different scientists of the instrument indication of the single instrument *I* requires each scientist to interact with a different part of *I*'s environment as different photons or phonons interact with their sense organs.

I conclude that even though a quantum model of environmental decoherence does not itself represent the physical interactions involved in measuring an observable on a quantum system, its application to those interactions can justify a claim as to the outcome of that measurement, based on personal observation of the indication of an appropriate instrument.

## 5. Quantum metrology and objective representation

Despite its successful applications in quantum metrology, measurement remains a problematic concept in quantum theory. The Born rule is usually said to give the probabilities of alternative possible outcomes of the measurement of an observable. But John Bell (1989) argued forcefully that the word 'measurement' should be banned altogether in quantum mechanics, and that the concept of measurement should not appear in an exact formulation of some serious part of that theory. Many people still take quantum theory to suffer a quantum measurement problem afflicting its conceptual foundations.[2] In one way of describing it, the problem is that quantum theory itself is inconsistent with the assumption that a competent measurement of an observable has an outcome. Even if that assumption is true of all such measurements that have been performed, recent arguments based on an extension of the scenario of Wigner's friend provide strong reasons to deny that every competent quantum measurement would have one and only one outcome in such a scenario.[3]

These problems challenge the assumption that a competent measurement of a quantum observable always *has* an absolute outcome. Desert pragmatism replaces this assumption with the view that a competent measurement of a quantum observable measurement has an outcome that is relative to a decoherence environment. But if a measurement does not have an absolute outcome that says what its value is, then what is the point of measuring a quantum observable, and how could the success of quantum metrology depend on such measurements? To answer these and related questions, we need a better understanding of quantum theory and how it is applied, in quantum metrology and more generally. That is what desert pragmatism offers.

The previous section explained how applications of quantum theory give us objective knowledge of measurement outcomes, quantum states and Born probabilities even though these are all relative to a physical situation offering the perspective of a hypothetical agent applying the theory. Here I will show how this explains why measurement of quantum observables is important, in quantum metrology and elsewhere.

The measurement of quantum observables is an important source of knowledge of the probabilities prescribed by the Born rule and of the quantum states to which the rule is generally taken to apply. But to exactly what does the Born rule assign probabilities? These are usually called (actual or possible) measurement outcomes. But if Bell was right, we should call them something else because the term 'measurement' should not appear in an exact formulation of any serious part of quantum mechanics. QBists call them *experiences* of an agent applying quantum mechanics. But if experiences are private to the agent who has them then their statistics cannot provide objective evidence for the quantum theory whose

---

[2] The problem dates back to the 1920's and the literature on it is too vast to cite.
[3] See Wigner(1962); Schmid, Ying, and Leifer (2023).

Born rule specifies their probabilities. Moreover, an automated measurement may be said to have occurred even before anyone becomes aware of its outcome (as in the case of an unmanned radar speed-trap).

According to desert pragmatism, the Born rule assigns probabilities to alternative meaningful claims $M_\sigma \varepsilon \Delta_j$ about the value of an observable $M$ on a system $\sigma$ in a suitable decoherence environment following an interaction between $\sigma$ and another system $A$ that results in $A$ acquiring a specific property $P{=}p_i$. $A$'s acquisition of this property is the physical event whose occurrence warrants magnitude claim $M_\sigma \varepsilon \Delta_i$ about $\sigma$. A magnitude claim about a system may be meaningful and true whether or not it concerns an event in which that magnitude is measured on that system. Not all true magnitude claims are measurement outcomes, but every outcome of a measurement of a quantum observable is based on a magnitude claim that is true relative to its decoherence environment.

Quantum metrology involves the application of quantum theory in metrology to enhance the precision with which physical quantities can be measured. These physical quantities include intervals of time and distance, frequency, current, voltage, acceleration, variations of magnetic and gravitational fields and of temperature. Note that all these quantities are classical—their values are part of the subject matter of classical (non-quantum) physics. Some correspond to observables in a quantum theory, but others do not. It makes sense to speak of quantum enhancement of the precision with which a physical quantity is measured only if that quantity is part of the subject matter of non-quantum physics. So even though one can use quantum-metrological techniques to measure features of quantum states (such as entanglement and purity) these do not enhance the precision with which non-quantum physical quantities are measured.

Some classical physical quantities do correspond to quantum observables represented by operators, while others do not. Electric and magnetic fields provide interesting examples, since their components are represented by "parameters" (scalar functions of position and time multiplying the identity operator) in the wave-equation of non-relativistic quantum mechanics, but by operators in quantum field theory. Non-relativistic quantum mechanics can be applied to enhance the precision of magnetic field measurements by estimating the phase of a quantum state whose unitary evolution is described by a Schrödinger equation in which the wave-function's phase is a function of classical magnetic field parameters.

In this and similar applications of quantum theory to metrology, a target quantity (here, magnetic field component) is not represented as a quantum observable to be measured directly. Instead, multiple measurements of *other* observables are performed on the quantum states of various probe systems whose phase is affected by interaction with the target quantity. The outcomes of these measurements enable one to estimate the value of the target quantity through its effect on the phase of the quantum state of the probe system(s). In this and many other instances of enhanced precision measurement in quantum metrology, measurement of quantum observables is performed not to find their values, but as a way of measuring a quantum state, and especially its phase. Whether or not a quantum measurement of one of these observables has an absolute outcome that reveals its true value is of no importance here. All that matters is that the statistics of these other measurement outcomes provide sufficient information about the objective quantum state of the probe system (relative to a relevant agent-situation) to permit a reliable inference to the result of measuring the value of the target quantity.

Such an application of quantum theory to enhance the precision of measurement of a non-quantum quantity bears on a dispute in the philosophy of metrology concerning the significance of the outcome of a measurement of that quantity. A realist holds that the outcome of a precise measurement of the quantity is accurate to the extent that it veridically represents the true value of that quantity. The goal of the measurement is veridical representation, and the measurement outcome is objective if and only if it veridically represents the mind-independent reality of that true value. This realism about measurement aligns with the representational sense of objectivity adopted by Burge (2009). Coherentists about measurement hold a different view of the significance of the outcome of a measurement of a non-quantum quantity. To quote Tal (2017, 44)

> Measurement outcomes are predictors that have been 'objectified' through coherence, and measurement accuracy is a special case of predictive accuracy.

Objectivity is here understood non-representationally, as emerging from the norm-guided activities of a community of scientists.

> The objectivity of measurement outcomes turns out to be grounded on considerations of coherence among models of multiple measurement processes securing such coherence.

Coherence is achieved when such activities are successful in aligning the measurement outcomes claimed by different practitioners.

> The argument from coherence explains how it is possible to assess the reliability of measuring instruments despite the inaccessibility of 'true' quantity values. (*ibid*)

In a footnote Tal does not exclude the possibility that these outcomes are aligned on the true value of the quantity:

> Whether such true values exist is a separate question, which is independent of the argument advanced here. (*ibid*)

But the norm-guided activities of the scientific community institute a nonrepresentational type of objectivity that contributes to the growth of scientific knowledge even if there turns out to be no true value of the measured quantity.

Returning to the use of quantum theory to enhance the precision of measurement of magnetic field components, one can now see how this is possible even if these components *have* no true values. The outcome of a quantum-enhanced measurement of a non-quantum magnetic field component is a claim about its value. But this claim is not directly based on a measurement of a quantum observable (represented by a self-adjoint operator) corresponding to this field component. It is based on measurement of *other* quantum observables sufficient to estimate the phase of a quantum state whose evolution is parameterized by real-number-valued magnetic field components.

In a quantum field theory such as quantum electrodynamics, one can represent components of the magnetic field by field operators (as components of the curl of the quantized 4-vector potential in some gauge). Each component is then treated as a quantum observable, a measurement of which typically fails to reveal its value. A realist who took a measurement of  a magnetic field component to reveal its true value as represented in non-relativistic quantum mechanics could no longer maintain their view in this new theoretical context. Progress in quantum theory would have shown that a magnetic field component has no true value, after all. This illustrates how realism about metrology is implicitly committed to the view that an accurate measurement of a quantity that figures in our currently best theory reveals that quantity's true value. To serve as a foundation on which better theories may be built, metrology must remain free of that commitment.

Gravitational waves predicted by the non-quantum theory of general relativity were first detected by LIGO (the Laser Interferometric Gravitational-wave Observatory). LIGO and similar devices are now used to observe, and measure properties of, merging black holes and neutron stars. The device is a sophisticated development of the Michelson interferometer whose null result is now considered to have provided support for Einstein's special theory of relativity. The passage of a gravitational wave through the evacuated cavities of the interferometer alters the relative distance between a pair of mirrors in each cavity by a tiny amount. This change in relative distance produces a change in the phase difference between the laser light propagating in the two cavities, and that varying phase difference produces a changing intensity when the beams are combined that is measured by radio-frequency photodiodes placed in the superposed beam.

Recent improvements to LIGO use quantum metrological techniques to enhance the sensitivity of the device by adding squeezed light into the interfering laser beam. Applying the quantum field theory of light, the superposed beam is initially modelled as a coherent quantum state whose standard phase-space representation has the form of a circular disc corresponding to equal indeterminacies in the position and momentum quadratures. After introduction of squeezed vacuum light (represented by an ellipse centred on the origin), the amplitude of the resulting squeezed light is less (respectively, more) indeterminate, while its phase indeterminacy is increased (respectively, decreased). The radio-frequency photodiodes respond to variations in amplitude and phase as a gravitational wave transits the interferometer by producing an electric current proportional to the corresponding intensity. Personal observation of records of instrument readings indicating this varying current is finally taken to constitute an observation of a gravitational wave: these instrument indications are taken as outcomes of measurements of variations in the relative distance between the pairs of mirrors in the two interferometer cavities.

From the point of view of metrology, the output of LIGO is a continuous sequence of measurement outcomes of the relative distances between two pairs of objects. This target quantity is non-quantum, and it is not represented as a quantum observable in an analysis of the experiment. Quantum theory is applied in engineering the device and analysing its operation. Specifically, the quantum theory of light is applied to understand and control the laser beams on which its operation depends. In a typical description of this application, the quantum state of the light is treated as something that can be prepared, manipulated and controlled, not as an idealization of an agent's degrees of belief about their own possible future experiences. But preparation, manipulation and control all involve causal interactions, so this description is also in tension with a pragmatist view that a quantum state is not a physical thing.

A pragmatist can resolve this tension by insisting that although to apply quantum field theory to light one presents a model that includes its quantum state, that model is not used to represent the physical state of the light but to offer good advice to a situated agent about what can be meaningfully said about possible future events involving the light and what degree of belief to hold about each. Light is a physical thing, but the correct quantum state to assign to it (relative to an agent-situation) is not. Preparing, manipulating and controlling light does not involve causal interaction with its quantum state.

Quantum theory is needed to understand how a radio-frequency photodiode works. But this understanding is provided not by quantum field theory, but by the simpler quantum theory of the photoelectric effect and the quantum theory of semiconductors. The photodiode measures the intensity of incident light by using energy from the light to free electron-hole pairs in the semiconductor. Classical physics may then be used to understand how the flow of these charge carriers through the photodiode generates a current whose magnitude is correlated with the intensity of the incident light.

To justify a claim about the measurement outcomes from LIGO one must appeal to a model of the measurement process. But what is a model, and how does modelling the measurement process provide the necessary justification? I quote Tal once more:

> By 'model' I mean an abstract and approximate representation of a local phenomenon, a representation that is used to predict (and sometimes also explain) aspects of that phenomenon. By 'modelling' I refer not only to the act of constructing models, but also to the iterative process of acquiring background knowledge, extracting predictions from a model, testing those predictions empirically, and modifying both the model and the concrete system it represents to achieve a better fit. (*op. cit.*, 33)

Two things are notable about this passage. Tal assumes that a model functions as a representation of a local phenomenon like (part of) a measurement process; and he acknowledges the importance of knowledge derived from a variety of external sources in modelling a process designed to measure a magnitude that figures in a theory.

Advanced LIGO was designed and modelled as a concrete realization of a measurement process whose output is a continuous sequence of measurement outcomes of the relative distances between two pairs of mirrors. This involved the use of models of the operation of many different elements of the whole apparatus, including the mirrors, interferometer cavities, laser light, and radiofrequency photo-diodes. Many different theories were applied in constructing these models, some quantum and others non-quantum (classical). The modelling process also appealed to background knowledge about the behaviour of all the elements of the apparatus derived not just from theory but from experiment.

An alternative use of the term 'model' is central to the semantic conception of a scientific theory that has come to be favoured by philosophers of science. This was proposed in conscious opposition to what came to be known as the syntactic conception of theories as deductively closed sets of sentences in some natural or formal language, itself a descendant of the (Euclidean, Newtonian) idea that a theory was whatever propositions followed logically from an initial set of axioms that stated its basic laws. The semantic conception, on the other hand, associates a theory with a set of mathematical structures called models which can be specified in any language, and which can be used to represent physical systems as having the same or a similar structure to one of its models. In this conception, a model is something that is associated with a particular theory, not something that may be independent of, or cobbled together from, applications of multiple, even incompatible, theories. It is important to recall this distinction of usage if one is to understand how quantum theory is applied in modelling the process of measurement in LIGO.

No single theory can be used to satisfactorily model the process of measurement in LIGO. In some sense quantum theories are more fundamental than non-quantum theories. But no quantum theory by itself can be used to model that entire process, and this is not required to justify accepting its measurement outcomes as objective knowledge. Focus now on how a particular quantum theory is used to model the laser light in LIGO. The lasers are said to produce 750 kW of power in the interferometer cavities. If the light were modelled using a classical theory of light, this would correspond to a determinate amplitude proportional to the square root of the power. But in a quantum theory, the laser light has no determinate amplitude, phase, or power—each is subject to quantum indeterminacy. Frequency-dependent squeezing of the light can be used to reduce the indeterminacy of amplitude while increasing the indeterminacy of phase, or vice versa. After interference and extraction from the beam, light power of the order of milliwatts is incident on the photo-diodes.

Naively, one might think that fluctuations in the current produced by a photo-diode correspond to fluctuations in the rate at which photons are incident on the device. But the light contains no determinate number of photons, just as it has no determinate phase or amplitude; and the quantum theory of light does not represent individual photons incident on the device. The quantum model of the light applies the Born rule to the objective quantum state of the light to specify an objective probability of a decoherence event in the photo-diode that can meaningfully and truly described as the production of an electron-hole pair in the semiconductor. With a high enough rate of such detections, the rate of pair productions in each short interval of time will almost certainly match the Born probability of their occurrence over that interval.

A model of the quantum theory of light is not applied to represent individual occurrences in the photo-diode or anywhere else. It is used to provide good advice to a user of the theory about the objective probability of detection events, and hence what degree of belief to hold about different relative frequencies of these events. The electric current produced by the photo-diode in effect rapidly measures these relative frequencies, permitting a reliable inference as to the quantum state of the light, and indirectly to the measurement outcomes that LIGO is taken to produce. This is a good example of how a quantum model can contribute to the justification of an inference to a measurement outcome from an instrument indication even though that model is not applied *to represent* any part of the physical process that connects them.

## 6. Conclusion

As I write, physicists are celebrating the 100th anniversary of the birth (in Heisenberg's seminal publication) of what is widely, and correctly, believed to be our most successful theory. But we are further than ever from a consensus on how this theory should be understood. Here is the headline of a news feature recently published in a prestigious journal:

> Physicists disagree wildly on what quantum mechanics says about reality, *Nature* survey shows. (*Nature* **643**, 1175-1179 (2025))

In fact, those surveyed were not asked their opinion on what quantum mechanics says about reality. Instead, they were asked "Which of the following, in your opinion, provides the best interpretation of quantum phenomena and interactions?", accompanied by a list of options, including the options "None of these (unknown)", "No need for an interpretation", "Other—non-categorized".

The headline of a news item often fails to accurately summarize its content, but here we see considerable confusion about what is meant by an interpretation, and how interpretation is connected to theories, phenomena and understanding. In his book offering an empiricist view of quantum mechanics, Van Fraassen (1991, 9) was careful to say what he meant by an interpretation of a scientific theory such as quantum mechanics:

> An interpretation of a theory is an answer to the question ‘How could the world possibly be how this theory says it is?’

In this view, an interpretation of quantum mechanics makes clear what quantum mechanics says about physical reality.

Desert pragmatism is not an interpretation of quantum mechanics in this sense since, according to a desert pragmatist, quantum theory says nothing about physical reality. Is this an admission that no-one understands quantum mechanics, as Feynman once famously claimed? No: it is rather the claim that to understand quantum mechanics one should explain how it is applied so successfully to a physical world it does not itself describe or represent.

The application of the quantum theory of light to improve the sensitivity of LIGO provides a striking example of how this works. LIGO was conceived and born to provide a novel means of epistemic access to the existence, properties and behaviour of distant and often ancient astrophysical objects and events. Its successful operation produces a variety of different representations of these things—not only linguistic descriptions, but also maps, graphs and parameter values in mathematical models (such as the masses of merged black holes). But none of this is represented by the quantum theory of light: that theory is used simply to make more precise measurements of the relative distances between two pairs of objects in the instrument itself. Here and elsewhere quantum theory is applied to improve and extend existing non-quantum representational resources. It is not used to provide novel quantum representational resources. This revelation may disappoint metaphysically-inclined philosophers and physicists who crave theoretical knowledge of ultimate physical reality on which to ground, or to which to reduce, all of physics. But it dissolves the paradox that after physicists have applied quantum theory successfully to a vast range of physical systems for more than a hundred years, not even they can agree on what it says about physical reality.

**References**

Adlam, E. and Rovelli, C. 2023 “Information is physical: Cross-perspective links in Relational Quantum Mechanics”, *Philosophy of Physics* **1** (1), 1–19.

Bell, J.S. 1989 “Against ‘measurement’”. In *62 Years of Uncertainty: Erice, 5–14 August 1989.* Plenum Publishers.

Brandom, R. 2000 *Articulating Reasons: An Introduction to Inferentialism* (Cambridge, MA: Harvard University Press).

Burge, T. 2009 “Perceptual Objectivity”, *Philosophical Review* **118** (3). 285–324.

Fine, A. 1982 “Joint distributions, quantum correlations, and commuting observables”, *Journal of Mathematical Physics* **23**, 1306–10.

Fuchs, C. 2023 “QBism: Where Next?” In Berghofer, P. and Wiltsche, H. (eds.) *Phenomenology and QBism* (Routledge, 2023), 78–143.

Healey, R. 2012 “Quantum Theory: a Pragmatist Approach”, *British Journal for the Philosophy of Science*, **63** (2012), 729–71.

Healey, R. 2022 “Securing the objectivity of relative facts in the quantum world”, *Foundations of Physics* **52**, 88.

Healey, R. 2026 *Pragmatism Works: Essays on Quantum Theory, Science and Metaphysics*. Oxford: Oxford University Press.

Kochen, S. and Specker, E. 1967 The problem of hidden variables in quantum mechanics. *Journal of Mathematics and Mechanics* **17**, 59-87.

Lewis, D. 1980 “A Subjectivist’s Guide to Objective Chance”, in Jeffrey, R. C. (ed.) *Studies in Inductive Logic and Probability*, *Volume II*. Berkeley: University of California Press, 263-93.

Muciño R., Okon, E. and Sudarsky, D. 2021. A reply to Rovelli's response to our “Assessing Relational Quantum Mechanics” arXiv preprint arXiv:2107.05817

Ollivier, H., Poulin, D. and Zurek, W. (2004) "Objective properties from subjective quantum states: environment as a witness", *Physical Review Letters* **93**, 220401.

Riedel, T. forthcoming “Is quantum relativism untameable? Revenge Wigner arguments for relative facts”, *British Journal for the Philosophy of Science* just accepted, manuscript available online.

Rovelli, C. 1996 “Relational Quantum Mechanics”, *International Journal of Theoretical Physics,* **35**,1637–78.

Schmid, D., Ying, Y. and Leifer, M. “A review and analysis of six extended Wigner’s friend arguments”, https://arxiv.org/abs/2308.16220

Sellars, W. 1953 “Inference and Meaning”, *Mind*, New Series, Vol. 62, No. 247, 313–338.

Tal, E. 2017 “Calibration: Modelling the Measurement Process”, *Studies in History and Philosophy of Science* **65–66**, 33–45.

Van Fraassen, B. 1991 *Quantum Mechanics: an Empiricist View*. Oxford: Clarendon Press

Wigner, E. 1962 “Remarks on the Mind–Body Question”, in I. J. Good (*ed*.), *The Scientist Speculates*, London: Heinemann.

Zurek, W. 2009 Quantum Darwinism. *Nature Physics* **5**, 181–188.